\documentclass[9pt]{article}
\usepackage{extsizes}
\usepackage[super,sort&compress,comma]{natbib} 
\usepackage[version=3]{mhchem}
\usepackage[left=1.5cm, right=1.5cm, top=1.785cm, bottom=2.0cm]{geometry}
\usepackage{balance}
\usepackage{times,mathptmx}
\usepackage{sectsty}
\usepackage{graphicx} 
\usepackage{lastpage}
\usepackage[format=plain,justification=raggedright,singlelinecheck=false,font={stretch=1.125,small,sf},labelfont=bf,labelsep=space]{caption}
\usepackage{float}
\usepackage{fancyhdr}
\usepackage{fnpos}
\usepackage[english]{babel}
\usepackage{array}
\usepackage{droidsans}
\usepackage{charter}
\usepackage[T1]{fontenc}
\usepackage[usenames,dvipsnames]{xcolor}
\usepackage{setspace}
\usepackage[compact]{titlesec}

\usepackage{epstopdf}
\definecolor{cream}{RGB}{222,217,201}

\usepackage{xcolor}
\usepackage[normalem]{ulem}

\newcommand{\change}[1]{#1}
\newcommand{\delete}[1]{}

\begin{document}

\pagestyle{fancy}
\thispagestyle{plain}
\fancypagestyle{plain}{

\renewcommand{\headrulewidth}{0pt}
}

\makeFNbottom
\makeatletter
\renewcommand\LARGE{\@setfontsize\LARGE{15pt}{17}}
\renewcommand\Large{\@setfontsize\Large{12pt}{14}}
\renewcommand\large{\@setfontsize\large{10pt}{12}}
\renewcommand\footnotesize{\@setfontsize\footnotesize{7pt}{10}}
\makeatother

\renewcommand{\thefootnote}{\fnsymbol{footnote}}
\renewcommand\footnoterule{\vspace*{1pt}%
\color{cream}\hrule width 3.5in height 0.4pt \color{black}\vspace*{5pt}} 
\setcounter{secnumdepth}{5}

\makeatletter 
\renewcommand\@biblabel[1]{#1}            
\renewcommand\@makefntext[1]%
{\noindent\makebox[0pt][r]{\@thefnmark\,}#1}
\makeatother 
\renewcommand{\figurename}{\small{Fig.}~}
\sectionfont{\sffamily\Large}
\subsectionfont{\normalsize}
\subsubsectionfont{\bf}
\setstretch{1.125} 
\setlength{\skip\footins}{0.8cm}
\setlength{\footnotesep}{0.25cm}
\setlength{\jot}{10pt}
\titlespacing*{\section}{0pt}{4pt}{4pt}
\titlespacing*{\subsection}{0pt}{15pt}{1pt}

\fancyfoot{}
\fancyfoot[RO]{\footnotesize{\sffamily{1--\pageref{LastPage} ~\textbar  \hspace{2pt}\thepage}}}
\fancyfoot[LE]{\footnotesize{\sffamily{\thepage~\textbar\hspace{3.45cm} 1--\pageref{LastPage}}}}
\fancyhead{}
\renewcommand{\headrulewidth}{0pt} 
\renewcommand{\footrulewidth}{0pt}
\setlength{\arrayrulewidth}{1pt}
\setlength{\columnsep}{6.5mm}
\setlength\bibsep{1pt}

\makeatletter 
\newlength{\figrulesep} 
\setlength{\figrulesep}{0.5\textfloatsep} 

\newcommand{\topfigrule}{\vspace*{-1pt}%
\noindent{\color{cream}\rule[-\figrulesep]{\columnwidth}{1.5pt}} }

\newcommand{\botfigrule}{\vspace*{-2pt}%
\noindent{\color{cream}\rule[\figrulesep]{\columnwidth}{1.5pt}} }

\newcommand{\dblfigrule}{\vspace*{-1pt}%
\noindent{\color{cream}\rule[-\figrulesep]{\textwidth}{1.5pt}} }

\makeatother


\sffamily
\begin{tabular}{m{4.5cm} p{11.5cm} }

DOI:10.1039/c6cp07919a & \noindent\LARGE{\textbf{Ab Initio Molecular Dynamics Relaxation and Intersystem Crossing Mechanisms of 5-Azacytosine$^\dag$ $^\bot$}} \\
\vspace{0.3cm} & \vspace{0.3cm} \\

 & \noindent\large Antonio Carlos Borin,$^{*\ddag}$\textit{$^{a}$} Sebastian Mai,$^{*\ddag}$\textit{$^{b}$} Philipp Marquetand,\textit{$^{b}$} and Leticia Gonz{\'{a}}lez\textit{$^{b}$} \\


 & \noindent\normalsize{
%
%
The gas phase relaxation dynamics of photoexcited 5-azacytosine has been investigated by means of SHARC (surface-hopping including arbitrary couplings) molecular dynamics, based on accurate multireference electronic structure computations.
Both singlet and triplet states were included in the simulations in order to investigate the different internal conversion and intersystem crossing pathways of this molecule.
It was found that after excitation, 5-azacytosine undergoes ultrafast relaxation to the electronic ground state with a time constant of about 1~picosecond.
Two important conical intersections have been identified as the funnel responsible for this deactivation mechanism.
The very low intersystem crossing yield of 5-azacytosine has been explained by the size of the relevant spin-orbit coupling matrix elements, which are significantly smaller than in related molecules like cytosine or 6-azauracil.
This difference is due to the fact that in 5-azacytosine the lowest singlet state is of $n_\mathrm{N}\pi^*$ nature, whereas in cytosine and 6-azauracil it is of $n_\mathrm{O}\pi^*$ character.
%
%
} \\

\end{tabular}



\renewcommand*\rmdefault{bch}\normalfont\upshape
\rmfamily
\section*{}
\vspace{-1cm}


\footnotetext{\textit{$^{a}$~Department of Fundamental Chemistry, Institute of Chemistry, University of S\~{a}o Paulo, NAP-PhotoTech the USP Consortium for Photochemical Technology, Av. Prof. Lineu Prestes 748, 05508-000, S\~{a}o Paulo, SP, Brazil. E-mail: Antonio Carlos Borin: ancborin@iq.usp.br}}

\footnotetext{\textit{$^{b}$~Institute of Theoretical Chemistry, Faculty of Chemistry, University of Vienna, W\"ahringer Stra\ss{}e 17, 1090 Vienna, Austria. E-mail: Sebastian Mai: sebastian.mai@univie.ac.at}}

\footnotetext{\dag~Electronic Supplementary Information (ESI) available: Optimized molecular geometries. See DOI: 10.1039/c6cp07919a}

\footnotetext{\ddag~These authors contributed equally to this work.}

\footnotetext{\textbf{$\bot$~This is the ArXiv version of the article: \textit{Phys. Chem. Chem. Phys.} \textbf{19}, 5888 (2017). DOI: 10.1039/c6cp07919a.}}



\section{Introduction}

\begin{figure}[b]
  \centering
  \includegraphics[scale=1]{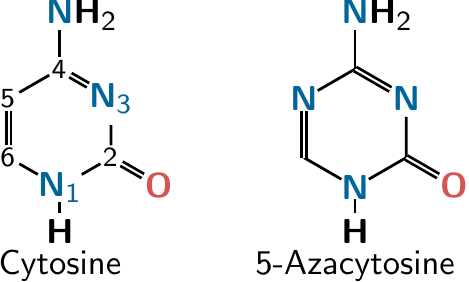}
  \caption{Schematic structures and atom numbering for cytosine and 5-azacytosine (5-AC).}
  \label{fig:Cyt-5AC}
\end{figure}

5-Azacytosine (5-AC) is an analogue of the nucleobase cytosine, obtained by the substitution of cytosine's C$_5$-H group by a nitrogen atom (see Figure~\ref{fig:Cyt-5AC}).
When this molecule is attached to a ribose sugar, its nucleotide \change{\emph{azacytidine}} is formed.
\change{Azacytidine} is an important chemotherapeutic drug,\cite{Lubbert2000CTMI,Cihak1974O} which acts by incorporation into DNA or RNA and by inhibiting methyltransferases.\cite{Santi1983C}
Of particular interest is the fact that chemotherapy using azabases and radiation/photodynamic therapies are often applied simultaneously,\cite{Kobayashi2009JPCA} a fact which draws attention to the photochemistry of 5-AC.
Furthermore, this molecule is also an interesting species for exploring the effect of chemical modifications of nucleobases on their photochemical properties, a topic which has received widespread attention.\cite{Kobayashi2009JPCA,Matsika2014TCC,Pollum2014TCC,Crespo-Hernandez2015JACS,Mai2016NC}

The excited states of 5-AC were investigated by means of steady-state absorption, transient absorption, and time-resolved thermal lensing experiments by Kobayashi et al.\cite{Kobayashi2009JPCA} 
They concluded that 5-AC is characterized by very low fluorescence, phosphorescence, and singlet oxygen yields.
These authors also noted that these properties are similar to those observed in 8-azaguanine, but differ strongly from 6-azauracil and 8-azaadenine, both of which show high singlet oxygen yields.
These differences were attributed to the state characters of the lowest singlet excited states; in particular, for 5-AC they reported that the two lowest singlet states were of $\pi\pi^*$ character, but a low-lying $n\pi^*$ state would be necessary for efficient intersystem crossing (ISC).\cite{Kobayashi2009JPCA} 

More recently, the excited-state potential energy surfaces (PESs) of 5-AC were investigated by one of us\cite{Giussani2014JCTC} employing the well-established CASPT2//CASSCF protocol,\cite{Merchan2013TCC} focusing on the possible relaxation and ISC mechanisms. 
Based on optimized excited-state minima/state crossings, minimum-energy paths, and linear interpolation scans, it was shown that the photochemistry of 5-AC differs from that observed in cytosine by the presence of a dark $^1n_\mathrm{N}\pi^*$ state below the bright $^1{\pi\pi^*}$ states, disagreeing with Kobayashi et al.\cite{Kobayashi2009JPCA}
Relaxation to the ground state was reported to be enabled by the presence of conical intersections (CoIns) involving the $^1n_\mathrm{N}\pi^*/S_0$ or $^1\pi\pi^*/S_0$ state pairs.
Low barriers to access these CoIns make ground state relaxation the most efficient photophysical process of 5-AC.
Singlet-triplet interaction regions were also identified, but it was still concluded that ISC should be only a minor side channel, in agreement with experimental findings.\cite{Kobayashi2009JPCA}

Since in many cases a photophysical reaction mechanism is controlled not only by the shape of PESs, but also by the details of the nuclear dynamics occurring on these surfaces, it is necessary to explicitly simulate this dynamics, considering all relevant nonadiabatic and spin-orbit couplings (SOCs).
Such nonadiabatic dynamics simulations allow obtaining excited-state lifetimes and identifying the actual branching ratios between the different possible relaxation pathways.
In this contribution, we present the first excited-state dynamics simulations of 5-AC, employing accurate multireference electronic structure methods for describing the PESs of all relevant states.
In order to investigate the ISC yields of 5-AC, we include both singlet and triplet states in the dynamics by means of the SHARC (surface hopping\cite{Tully1990JCP} including arbitrary couplings) method.\cite{Richter2011JCTC,Mai2015IJQC}
Together with the static picture of the relaxation mechanisms described previously,\cite{Giussani2014JCTC} we contribute further insights into the relaxation process of this important aza-compound.


\section{Computational details}

Energies, analytical gradients \cite{Shepard1992JCP,Shepard1995WorldScientific,Lischka2002MP}, non-adiabatic couplings (for optimizations of CoIns)\cite{Lischka2004JCP,Dallos2004JCP} and spin-orbit couplings\cite{Mai2014JCP_reindex} were calculated using the MRCIS(6,5)/def2-SVP (multireference configuration interaction including single excitations) level of theory, provided by the COLUMBUS 7.0\cite{Lischka2011WCMS,Lischka2012a,Yabushita1999JPCA,Mai2014JCP_reindex} quantum chemistry package.
All atomic and spin-orbit mean-field integrals were taken from MOLCAS\cite{Aquilante2010JCC} through the MOLCAS-COLUMBUS interface.\cite{Lischka2011WCMS}
The singlet-and-triplet excited-state dynamics of 5-AC was simulated using the development version of the SHARC ab initio dynamics package.\cite{Richter2011JCTC,Mai2015IJQC,Mai2014SHARC}
To avoid computing nonadiabatic coupling vectors during the dynamics, the wavefunction propagation was carried out with the local diabatization procedure,\cite{Granucci2001JCP} where the required wavefunction overlaps were computed with our recent overlap code.\cite{Plasser2016JCTC}


\subsection{Ab initio level of theory}\label{sec:abinitio}

\begin{figure}
  \centering
  \includegraphics[scale=1]{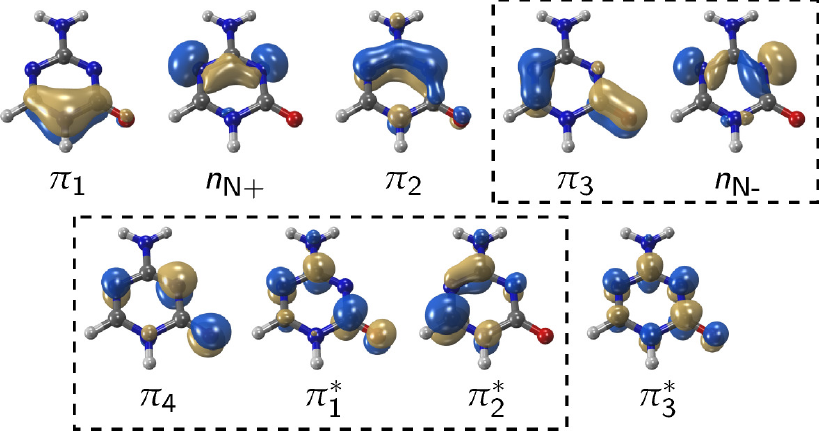}
  \caption{Active space of 5-AC in terms of state-averaged natural orbitals, computed at the SA(3S+2T)-CASSCF(12,9) level, in the ground-state equilibrium geometry and with the def2-svp basis sets.
  The orbitals enclosed by dotted lines constitute the CAS(6,5) reference space for the MRCIS calculation.}
  \label{fig:5AC-Act-Space}
\end{figure}

Stationary points, CoIns, and singlet-triplet minimum-energy crossings were optimized at the MRCIS level of theory with the def2-SVP basis set,\cite{Weigend2005PCCP} using a CAS(6,5) reference space.
The orbitals were taken from a SA(3S+2T)-CASSCF(12,9) calculation (complete-active-space self-consistent-field\cite{Roos1987ACP}), where a single set of orbitals is optimized for the average over 3~singlet and 2~triplet states.
At the ground-state optimized geometry, these orbitals are the two $n_\mathrm{N}$ and 7 $\pi$ orbitals shown in Figure~\ref{fig:5AC-Act-Space}; the subset of orbitals included in the reference space of the MRCIS(6,5) calculation is highlighted in the same figure.
The $n_{\mathrm{O}}$ orbital was kept inactive because the $n_{\mathrm{O}}\pi^*$ state is too high in energy to take part in the main events discussed in this work.\cite{Giussani2014JCTC}
Inner shell orbitals were frozen in the CI step of the computation and the Douglas-Kroll-Hess scalar-relativistic Hamiltonian \cite{Douglas1974AP,Hess1986PRA,Reiher2012WCMS} was employed.
This level of theory is abbreviated as ``MRCIS(6,5)/def2-SVP'' in the following and is used consistently throughout the manuscript.

\begin{table*}
  \centering
  \caption{Computed energies (eV) for the most relevant singlet and triplet states of 5-AC at different important geometries, from previous CASPT2//CASSCF computations\cite{Giussani2014JCTC} and from the MRCIS(6,5)/def2-SVP computations of this work.
  All energies are relative to the ground state minimum energy and energies of the optimized states are given in bold.}
  \label{Tab:en-mrcis-2}
  \begin{tabular}{l ccccc c ccc c cccc}
    \hline 
            & \multicolumn{5}{c}{\change{SS-}CASPT2(14,10)/ANO-L (from Ref.~\cite{Giussani2014JCTC})} && \multicolumn{3}{c}{\change{MS-CASPT2(14,10)/ANO-L}} && \multicolumn{4}{c}{MRCIS(6,5)/def2-SVP}\\
    \cline{2-6} \cline{8-10} \cline{12-15}
    Structure                                       &$S_0$           &$^1n_\mathrm{N}\pi^*$ &$^1\pi\pi^*$    &$^1n_\mathrm{O}\pi^*$ &$^3\pi\pi^*$     &&$S_0$                 &$^1n_\mathrm{N}\pi^*$    &$^1\pi\pi^*$       &&$S_0$                 &$^1n_\mathrm{N}\pi^*$    &$^1\pi\pi^*$     &$^3\pi\pi^*$ \\
    \hline
    $S_0$ min                                       &$\mathbf{0.00}$ &$4.46$                &$4.74$          &$5.58$                &$4.33$           &&\change{$\mathbf{0.00}$} &\change{$4.61$} &\change{$5.06$}          &&$\mathbf{0.00}$ $^a$  &$5.07$                   &$5.31$           &$4.62$ \\
    $^1n_\mathrm{N}\pi^*$ min                       &$1.52$          &$\mathbf{3.59}$       &$4.38$          &---                   &$3.71$           &&--- &--- &---                                                       &&$1.30$                &$\mathbf{4.10}$          &$5.76$           &$4.28$ \\
    $^1\pi\pi^*$ min                                &$1.39$          &$4.67$                &$\mathbf{3.69}$ &$4.08$                &$3.57$           &&--- &--- &---                                                       &&$1.22$                &$4.90$                   &$\mathbf{4.28}$  &$3.82$ \\
    $^3\pi\pi^*$ min                                &$1.06$          &$4.02$                &$3.76$          &$4.65$                &$\mathbf{3.27}$  &&--- &--- &---                                                       &&$1.50$                &$4.32$                   &$5.35$           &$\mathbf{3.54}$ \\
    $^1\pi\pi^*/{}^1n_\mathrm{N}\pi^*$ CoIn         &$0.08$          &$\mathbf{3.97}$       &$\mathbf{4.04}$ &$4.80$                &$3.63$           &&--- &--- &---                                                       &&$0.87$                &$\mathbf{4.41}$          &$\mathbf{4.41}$  &$3.78$ \\
    $S_1/S_0$ CoIn ($\alpha$)                       &$\mathbf{3.75}$ &---                   &$\mathbf{3.87}$ &$3.76$                &$4.27$           &&\change{$\mathbf{4.82}$ $^b$} &\change{$5.15$} &\change{$\mathbf{4.83}$ $^b$}                   &&\change{$\mathbf{5.31}$ $^b$} &\change{$7.03$} &\change{$\mathbf{5.31}$ $^b$}              &---    \\
    $S_1/S_0$ CoIn ($\beta$)                        &$\mathbf{3.47}$ &$5.12$                &$\mathbf{3.47}$ &---                   &$3.89$           &&\change{$\mathbf{3.76}$} &\change{$5.66$} &\change{$\mathbf{3.77}$}                   &&$\mathbf{4.06}$       &$6.73$                   &$\mathbf{4.06}$  &$4.01$ \\
    $S_1/S_0$ CoIn ($\gamma$)                       &---             &---                   &---             &---                   &---              &&\change{$\mathbf{3.84}$} &\change{$\mathbf{3.85}$} &\change{$5.90$}                   &&$\mathbf{3.99}$       &$\mathbf{3.99}$          &$6.21$           &$5.95$ \\
    $^1\pi\pi^*/^3\pi\pi^*$ STC                     &$3.89$          &---                   &$\mathbf{3.87}$ &---                   &$\mathbf{3.77}$  &&--- &--- &---                                                       &&---                   &---                      &---              &---    \\
    $^1n_\mathrm{N}\pi^*/^3\pi\pi^*$ STC            &---             &---                   &---             &---                   &---              &&--- &--- &---                                                       &&$1.19$                &$\mathbf{4.12}$          &$5.57$           &$\mathbf{4.12}$ \\
    \hline
  \end{tabular}

  $^a$ MRCIS(6,5)/def2-SVP ground state minimum energy: --408.6531755~Hartree.

  \change{$^b$ Optimized with constrained C$_2$=O bond length.}
\end{table*}

\subsection{Dynamics}

\begin{figure}
  \centering
  \includegraphics[scale=1]{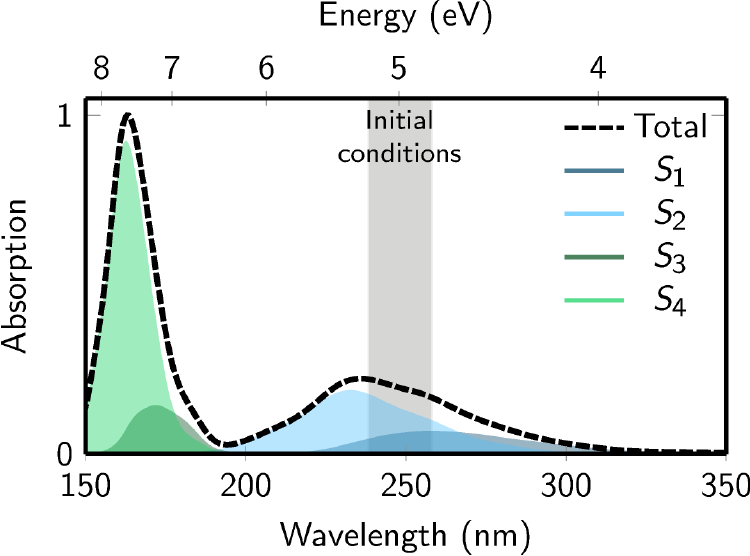}
  \caption{Simulated absorption spectrum for 5-AC at the MRCIS(6,5)/def2-SVP level of theory, based on 1000~geometries.
  The plot also shows the individual contributions of the excited states $S_1$ to $S_4$ and the energy range from where initial conditions for the dynamics were selected.
  }
  \label{fig:spectrum}
\end{figure}

The generation of initial conditions for the dynamics simulation was based on an optimization of the ground-state equilibrium geometry, followed by the computation of the harmonic frequencies.
From these frequencies, a quantum harmonic oscillator Wigner distribution\cite{Dahl1988JCP,Schinke1995} of the lowest vibrational state was computed, from which 1000 uncorrelated geometries and velocities were sampled.
For each sampled geometry, a vertical excitation calculation including 5~singlet states was carried out.
An absorption spectrum was constructed as a superposition of Gaussian functions, centered on the computed vertical excitation energies and with the height proportional to the oscillator strength; the spectrum is presented in Figure~\ref{fig:spectrum}.
For dynamics, initial conditions were stochastically chosen\cite{Barbatti2007JPPA} from the first absorption band in the energy window 4.8--5.2~eV, yielding 146 initial conditions (47 starting in the $S_1$ state and 99 in the $S_2$ state).
\change{The excitation window was chosen to be slightly below the absorption maximum, as is commonly done in many experiments on nucleobases,\cite{Pollum2014TCC} with a width such that we obtain a sufficient number of initial conditions.}

The dynamics simulation considered 3~singlet and 2~triplet states, where all triplet components were explicitly included (yielding 9~states in total).
The trajectories were propagated until 1000~fs or until relaxation to ground-state occurred, with a nuclear time step of 0.5~fs and an electronic time step of 0.02~fs.

\subsection{\change{CoIn optimizations}}

\change{
Optimizations of the $S_1/S_0$ CoIns were also carried out with the MS-CASPT2 method, using numerical gradients, the algorithm of Levine et al.,\cite{Levine2008JPCB} and the \textsc{Orca} optimizer.\cite{Neese2012WCMS}
The calculations employed orbitals from an SA(4S)-CASSCF(14,10)/ANO-L-VDZP calculation with Cholesky decomposition and a non-relativistic Hamiltonian; in the MS-CASPT2 step, all four roots were included, the IPEA shift\cite{Ghigo2004CPL} was set to zero, and an imaginary shift\cite{Forsberg1997CPL} of 0.2~a.u.\ was applied.
}


\section{Results and discussion}


\subsection{Validation of the level of theory}

\begin{figure}
  \centering
  \includegraphics[scale=1]{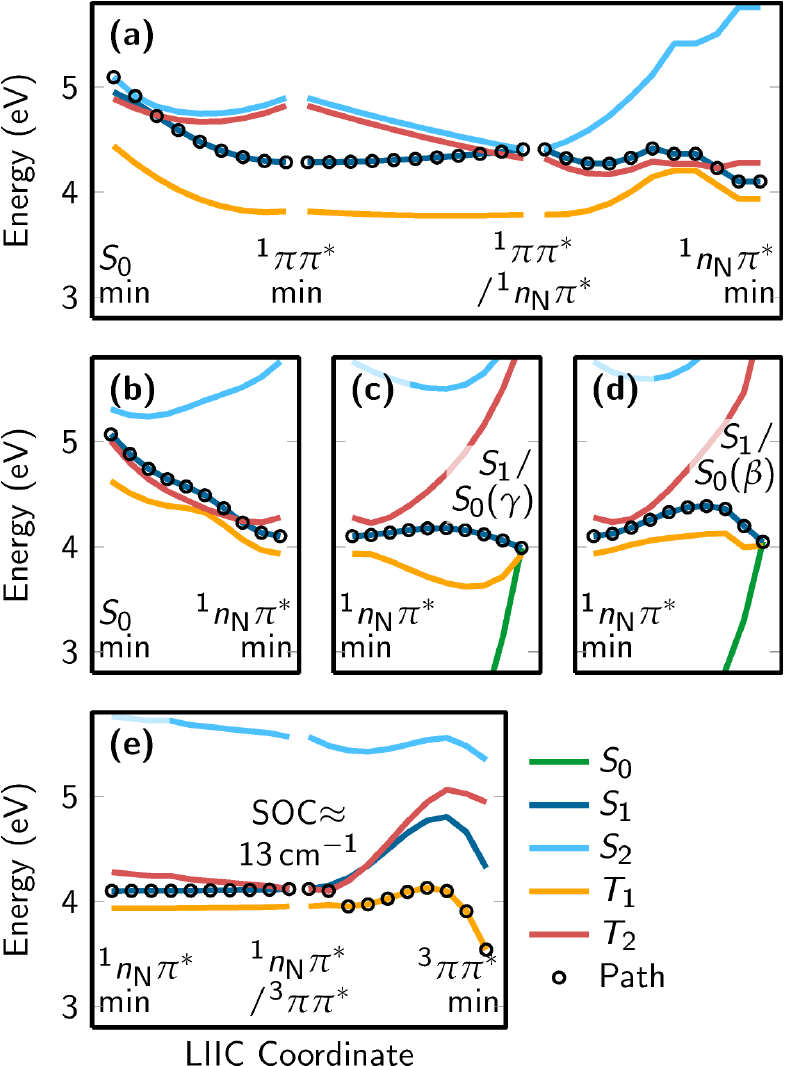}
  \caption{Linear interpolation in internal coordinates (LIIC) scans connecting important optimized geometries (as indicated by the labels).
  Computed at MRCIS(6,5)/def2-SVP.
  }
  \label{fig:liic}
\end{figure}

The static picture of the photochemistry of 5-AC was investigated previously,\cite{Giussani2014JCTC} employing the photochemical reaction approach and the CASPT2//CASSCF protocol.
Due to the high expense of dynamics simulations, here we describe the excited-state PESs with a more cost-efficient electronic structure method than CASPT2, namely the MRCIS method, as detailed above.
To validate that MRCIS can provide accurate (with respect to the previous CASPT2/CASSCF results) PESs for the dynamics simulations, we carried out a series of test calculations aimed at reproducing the fundamental characteristics of the PESs revealed with the CASPT2//CASSCF protocol.\cite{Giussani2014JCTC}
In order to do that, the geometries of all relevant structures (excited-state minima, CoIns, singlet-triplet crossings (STC)) for the excited state dynamics of 5-AC were reoptimized at the MRCIS level.
Subsequently, excitation energies were computed and compared with the benchmark results obtained with the CASPT2//CASSCF protocol \cite{Giussani2014JCTC}.
The results are collected in Table~\ref{Tab:en-mrcis-2}; all optimized geometries are given in the ESI.$^\dag$

In the Franck-Condon region, both levels of theory (MRCIS and CASPT2) agree that the lowest excited singlet state is of $^1n_\mathrm{N}\pi^*$ character, with the bright $^1\pi\pi^*$ state slightly (about 0.25~eV) above.
\change{The experimental absorption maxima are reported to be 4.86~eV\cite{Raksanyi1978B} (thin film on CaF$_2$) and 5.3~eV\cite{Kobayashi2009JPCA} (in acetonitrile), hence both MS-CASPT2 (5.06~eV) and MRCIS (5.31~eV) appear to correctly predict the vertical excitation energy of the bright $^1\pi\pi^*$ state.}
The lowest triplet state, $^3\pi\pi^*$, is predicted to be lower in energy than the lowest singlet state.
Outside the Franck-Condon region, at both levels of theory, three excited-state minima were obtained, for the $^1\pi\pi^*$, $^1n_\mathrm{N}\pi^*$ and $^3\pi\pi^*$ states. 
Furthermore, a crossing of $^1\pi\pi^*$ and $^1n_\mathrm{N}\pi^*$ states, and several crossings of the $S_0$ with the excited states were found.
Note that the wavefunction characters at the $S_1/S_0$ CoIns are strongly mixed and we cannot denote them as either ``$^1n_\mathrm{N}\pi^*/S_0$'' or ``$^1\pi\pi^*/S_0$''.
Instead, we simply refer to them as ``$S_1/S_0$ CoIn and label the different CoIns with $\alpha$, $\beta$, and $\gamma$.

In order to check for any relevant barriers between the mentioned geometries, we also performed linear interpolation in internal coordinates (LIIC) scans, as shown in Figure~\ref{fig:liic}.
The plots shown in panels (a) and (b) of this figure indicate that there are low-barrier paths from the Franck-Condon region (``$S_0$ min'') to the $^1n_\mathrm{N}\pi^*$ minimum, which is the lowest minimum on the $S_1$ adiabatic surface of 5-AC.
From there, several relaxation pathways exist, all associated with only small barriers.
\change{Note that these are not the true barriers, due to the fact that they were obtained with the LIIC scans.
The true barriers---which will be invariably smaller---could be obtained with minimum energy paths.
Nevertheless, the barriers will be larger than zero because a stable $^1n_\mathrm{N}\pi^*$ minimum was found.
}

Panels (c) and (d) present the paths from the $^1n_\mathrm{N}\pi^*$ minimum to CoIns between the ground state and the lowest excited singlet state.
Both CoIns are local minima on the $S_1/S_0$ intersection seam, separated by only small barriers (smaller than 0.07~eV and 0.28~eV, respectively) from the $^1n_\mathrm{N}\pi^*$ minimum.
The crossing in panel (c), $S_1/S_0$ CoIn $\gamma$, has not been reported previously as an $S_1/S_0$ CoIn, but it is related to the ``$(^3\pi\pi^*/\mathrm{gs})_\mathrm{STC}$'' geometry of Giussani et al.\cite{Giussani2014JCTC} (which is actually a $S_0/S_1/T_1$ three-state degeneracy point according to their energy profiles).
It is characterized by a pronounced boat-like conformation of the ring.
The crossing in panel (d) can be identified as the ``$(^1\pi\pi^*/\mathrm{gs})_{\mathrm{CI}-\beta}$'' geometry from these authors,\cite{Giussani2014JCTC} showing an envelope-like geometry where carbon C$_6$ is displaced out of the ring plane.

Note that the ``$(^1\pi\pi^*/\mathrm{gs})_{\mathrm{CI}-\alpha}$'' CoIn, the other $S_1/S_0$ CoIn reported previously,\cite{Giussani2014JCTC} can be easily distinguished by having a much longer C$_2$=O bond due to a different electronic character; here, at MRCIS level we did not obtain this CoIn.
\change{
Hence, we reoptimized the three CoIns from Ref.,\cite{Giussani2014JCTC} using MS-CASPT2 instead of SS-CASPT2,\cite{Giussani2014JCTC} because the latter does not correctly describe CoIns.\cite{Malrieu1995TCA}
The energies of the CoIns are given in Table~\ref{Tab:en-mrcis-2} and the geometries and further details are given in the ESI.$^\dag$
Interestingly, at the MS-CASPT2 level of theory the $S_1/S_0$ ($\alpha$) CoIn is not a minimum on the intersection seam, but the optimization converges towards $S_1/S_0$ ($\gamma$), the same behavior which is also observed at the MRCIS level of theory.
}

Finally, panel (e) of Figure~\ref{fig:liic} shows a path from the $^1n_\mathrm{N}\pi^*$ minimum to a minimum-energy crossing point with the $^3\pi\pi^*$ state and from there to the $^3\pi\pi^*$ minimum.
Because the barrier between the former two points is only 0.02~eV, ISC is in principle possible from the $^1n_\mathrm{N}\pi^*$ minimum.
However, the small SOC of only 13~cm$^{-1}$ probably does not allow for fast ISC.

Comparing the results obtained at the MRCIS level of theory with the previous \change{and new} CASPT2 results, we observe that MRCIS consistently predicts higher energies for all states (see Table~\ref{Tab:en-mrcis-2}).
However, the relative energies in the Franck-Condon region and among the critical points are well described at MRCIS level of theory; the largest difference is found in the energy of the $^3\pi\pi^*$ minimum (0.32~eV below the $^1n_\mathrm{N}\pi^*$ minimum with CASPT2 vs.\ 0.56~eV with MRCIS).
All other energies relative to the $^1n_\mathrm{N}\pi^*$ minimum agree within 0.1~eV between these methods.
Therefore, and because the computed LIIC scans computed at the MRCIS level of theory compare well with the scans from Ref.~\cite{Giussani2014JCTC}, we conclude that MRCIS correctly reproduces all key aspects of the PESs of 5-AC.


\subsection{Dynamics simulations}

\begin{figure}
  \centering
  \includegraphics[scale=1]{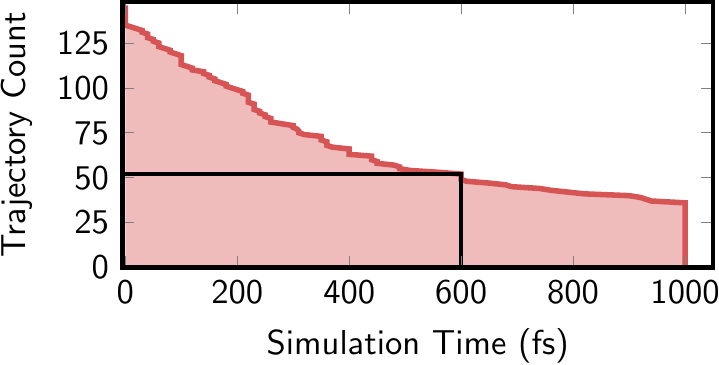}
  \caption{Number of trajectories which have at least a given simulation time.
  The black rectangle corresponds to the subset of 52~trajectories included in the presented analysis.
  }
  \label{fig:ntraj}
\end{figure}

Due to the fact that sometimes some active orbital had an occupation number very close to 2, a number of trajectories experienced convergence problems in the CASSCF computation at different time steps and terminate before reaching 1000~fs simulation time.
Figure~\ref{fig:ntraj} presents how many trajectories have achieved at least a given simulation time.
In order to find a good compromise between analyzed simulation time and number of trajectories included in the ensemble analysis, we considered all trajectories with at least 600~fs simulation time (black rectangle in Figure~\ref{fig:ntraj}), corresponding to 52~trajectories out of the initially started 146~ones.
This restriction is necessary to avoid an overestimation of the $S_1$ life time, which would occur if unreactive, early-terminated trajectories would be included.
The subset of 52~trajectories reproduces accurately the initial $S_1$/$S_2$ ratio: 32.7\% of the trajectories (17 of 52) started in $S_1$, which compares well with the 32.2\% ratio (47 of 146) found in the full set of initial conditions.
Furthermore, the initial geometries of the 52 selected trajectories show the same coordinate distribution as the excluded trajectories.
Hence, we are certain that the analysis of only 52~trajectories does not lead to a bias in the results.

\begin{figure}
  \centering
  \includegraphics[scale=1]{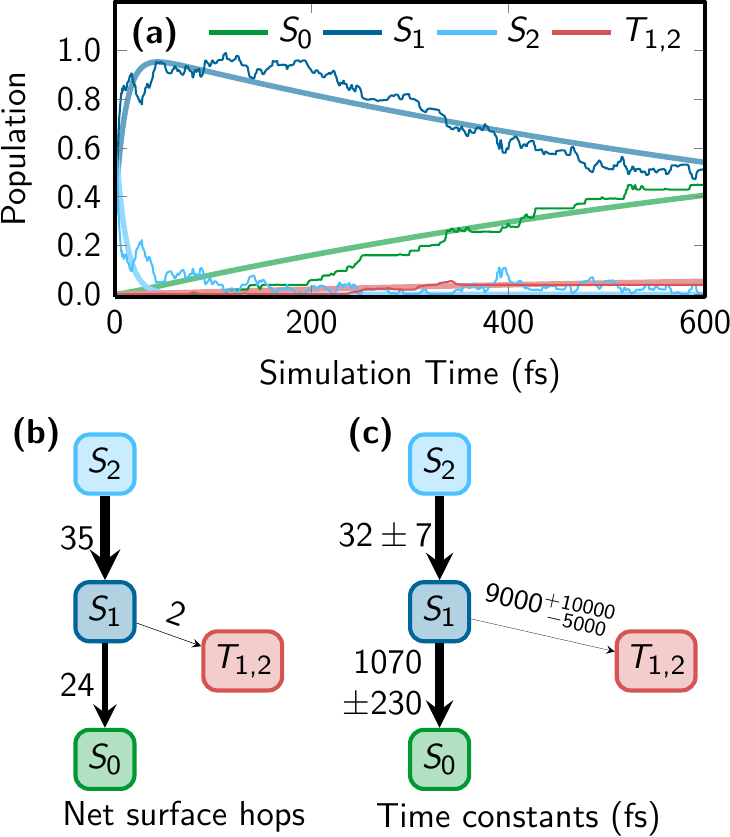}
  \caption{Results of the excited-state dynamics of 5AC, using SHARC and MRCIS(6,5)/def2-SVP.
  Panel (a) shows the evolution of the electronic populations, panel (b) the net surface hops between the electronic states, and panel (c) the globally fitted time constants including errors. 
  }
  \label{fig:pop}
\end{figure}

As the first part of the trajectory analysis, Figure~\ref{fig:pop} presents the temporal evolution of the excited-state populations (plotted in panel (a)).
Note that only populations in the adiabatic basis ($S_1$, $S_2$, ...) are presented, since the $n_\mathrm{N}\pi^*$ and $\pi\pi^*$ states mix strongly during the dynamics and a rigorous diabatization of the populations is not feasible.
It can be seen that, qualitatively, the system starts in a mixture of $S_1$ and $S_2$, but quickly evolves to populate the $S_1$ state with almost 100\%.
Subsequently, the $S_1$ population decreases, and the $S_0$ population increases, accompanied by a very small increase of the triplet population.

For a quantitative analysis of the population evolution, we used the net number of hops from the simulations (Figure~\ref{fig:pop} panel (b)) to decide on a kinetic reaction model, which is shown in Figure~\ref{fig:pop} (c).
From the model, consisting of four species ($S_0$, $S_1$, $S_2$, $T_{1,2}$) and three elementary reactions, it is possible to find the time laws for the species' populations, which where fit globally to the population data from the simulations.
This procedure yielded three time constants, whose statistical errors were computed with the bootstrapping method\cite{Nangia2004JCP} based on 1000~resamples of the ensemble.
Note that these errors describe the uncertainties related to the size of the ensemble and therefore allow judging whether the number of trajectories is sufficient.
The errors do not include the errors due to the choice of electronic structure and dynamics methods.

The first time constant, $\tau_1=32\pm 7$~fs, describes the internal conversion from $S_2$ to $S_1$.
Note that this is not identical to a $^1\pi\pi^*\rightarrow{}^1n_\mathrm{N}\pi^*$ time constant, because the adiabatic $S_1$ potential can have either $^1\pi\pi^*$ or $^1n_\mathrm{N}\pi^*$ character, depending on geometry.
In order to obtain an estimate for the $^1\pi\pi^*\rightarrow{}^1n_\mathrm{N}\pi^*$, we analyzed the ensemble average of the active state oscillator strength.
This value decreases by an order of magnitude within the first 40~fs; hence, the actual $^1\pi\pi^*\rightarrow{}^1n_\mathrm{N}\pi^*$ process is most likely occurring on this time scale.
The second time constant, with a value of $\tau_2=1070\pm 230$~fs, describes the relaxation from the $S_1$ PES to the electronic ground state.
Finally, the value of the third time constant is $\tau_3=9000\substack{+10000\\-5000}$~fs, corresponding to ISC from $S_1$ to the triplet states.
The associated statistical error is very large due to the fact that only two trajectories have sampled the ISC pathway.
Significantly more trajectories, and a longer simulation time, would be necessary to obtain a more precise estimate of this time constant.

Using the two latter time constants, it is possible to estimate the total triplet yield at long simulation times, given by $\Phi_\mathrm{ISC}=\tau_2/(\tau_2+\tau_3)$.
Including the statistical uncertainties of $\tau_2$ and $\tau_3$, we obtain an ISC yield of $10\pm 8$\%, where the error does not include the uncertainty due to the method choices, as mentioned above.
This value is higher than the experimental singlet oxygen yield of $<1$\% reported in the literature for 5-AC,\cite{Kobayashi2009JPCA} but it still shows clearly that ISC in 5-AC is only a minor side channel next to the much more efficient ground state relaxation.
It should also be noted that our simulations only consider gas-phase 5-AC, whereas the experiments were carried out in acetonitrile and it is well known that solvent effects can influence the ISC yield of nucleobases.\cite{Hare2006JPCB}
\change{For a comparison with the ISC yield of cytosine, see below.}

\begin{figure}
  \centering
  \includegraphics[scale=1]{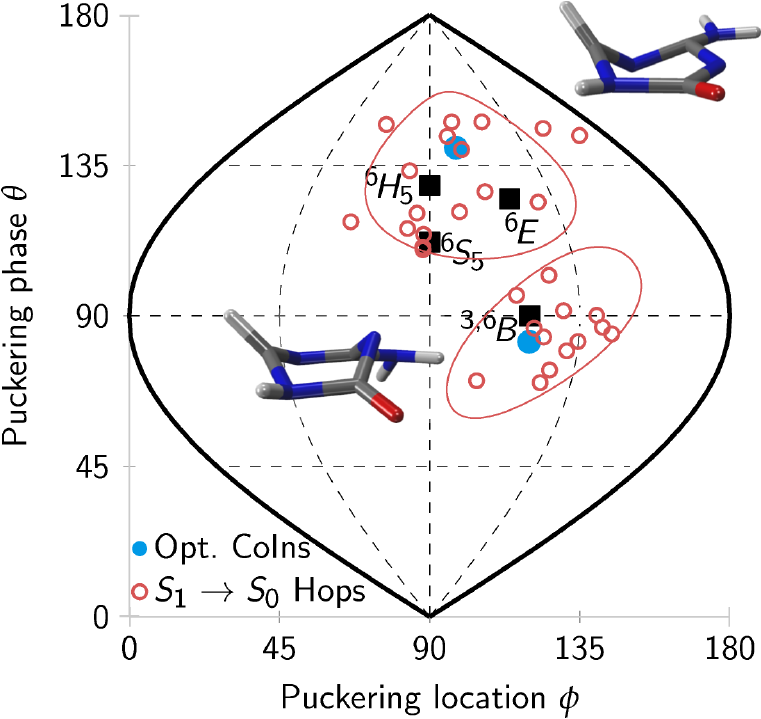}
  \caption{
  Location of the $S_1\rightarrow S_0$ hopping geometries (small open circles) in the space of the Cremer-Pople parameters $\phi$ and $\theta$,\cite{Cremer1975JACS} with the plot spanning the space of all possible conformations of 6-membered rings.
  The locations of the CoIn geometries optimized with MRCIS (filled circles; $S_1/S_0$ ($\beta$) on the top and $S_1/S_0$ ($\gamma$) in the middle) and of selected ring conformations (squares; boat $^{3,6}B$, half-chair $^6H_5$, screw-boat $^6S_5$, envelope $^6E$)\cite{Boeyens1976JCMS} are also indicated.
  }
  \label{fig:geoms}
\end{figure}

The nature of the ground state relaxation pathway was investigated by analyzing the geometries where surface hops from the $S_1$ to the $S_0$ state occurred.
For these 30~geometries (taken not only from the 52 longest trajectories, but from all 146), we computed the Cremer-Pople parameters, which characterize the ring conformation.\cite{Cremer1975JACS}
For a six-membered ring, there are three Cremer-Pople parameters, $Q$, $\phi$, and $\theta$; $Q$ is the puckering amplitude, and the two angles give the location and type of ring puckering.
For the $S_1\rightarrow S_0$ hopping geometries, we find an average $Q$ of 0.5\AA, which is significantly larger than the ensemble average of 0.2\AA, showing that in all cases, ground state decay involves a strongly deformed heteroaromatic ring.

The two puckering angles span the surface of a half-sphere, on which certain parameter combinations $(\phi,\theta)$ represent certain ring conformations like chair, boat, or envelope.\cite{Boeyens1976JCMS}
In Figure~\ref{fig:geoms}, we plot the location of all hopping geometries in this two-dimensional space.
It can be seen that all hopping geometries cluster around two relatively small regions in this conformation space.
One cluster is located close to the $^{3,6}B$ geometry, which is a boat where atoms N$_3$ and C$_6$ are displaced from the molecular plane in the same direction.
This geometry is very close to the $S_1/S_0$ ($\gamma$) CoIn from Table~\ref{Tab:en-mrcis-2} and Figure~\ref{fig:liic} (c).
The second cluster is close to the $^6H_5$ half-chair conformation, where atoms N$_5$ and C$_6$ are displaced from the ring plane in opposite directions, and to the envelope $^6E$ structure, where atom C$_6$ is displaced.
This cluster is also in the vicinity of the $S_1/S_0$ ($\beta$) CoIn reported above.
Because all hopping geometries show a C$_2$=O bond length of below 1.3\AA, we find that none of the trajectories employed the previously reported ``$(^1\pi\pi^*/\mathrm{gs})_{\mathrm{CI}-\alpha}$'' CoIn,\cite{Giussani2014JCTC} which has a C$_2$=O bond length of 1.4\AA.
Interestingly, all hopping geometries to some extent include a ring deformation localized close to the N$_5$=C$_6$ double bond of 5-AC, indicating that this bond is crucial for the ground state relaxation of this molecule.
In this respect, it would be interesting to investigate what effect substitution of the C$_6$-H group would have on the excited-state dynamics.

\subsection{Photophysics of 5-azacytosine}

The most likely relaxation pathway of 5-AC is, in terms of the PESs in Figure~\ref{fig:liic}, a sequence of panel (a) or (b)---depending on whether the molecule is excited to $S_1$ or $S_2$---followed by either panel (c) or (d).
According to Figure~\ref{fig:geoms}, both of the employed $S_1/S_0$ CoIns contribute approximately equally to ground state relaxation, with 17~hops in the $^6H_5/^6E$ region and 13~hops in the $^{3,6}B$ region.
These results also show that at our MRCIS level of theory, all trajectories relax via the $^1n_\mathrm{N}\pi^*$ state, the starting point in both Figure~\ref{fig:liic} (c) and (d).
Probably due to the short life time of the $^1\pi\pi^*$ population, no trajectory followed the pathway from the $^1\pi\pi^*$ minimum to the ``$(^1\pi\pi^*/\mathrm{gs})_{\mathrm{CI}-\alpha}$'' CoIn, even though this pathway would be almost barrierless according to Giussani et al.\cite{Giussani2014JCTC}

The trajectories also showed clearly that ISC in 5-AC is only a minor relaxation channel.
Actually, the ISC yield obtained here ($10\pm 8$\%) is much lower than the ISC yield of about 32\% reported for previous SHARC simulations of \change{gas-phase} cytosine, the parent nucleobase of 5-AC.\cite{Mai2013C}
\change{The low ISC yield of 5AC is not primarily caused by 5AC avoiding the singlet-triplet degeneracy region.
Actually, we found that the $S_1-T$ gap is small ($<$0.02~eV) in about 5\% of all time steps in the trajectories and the $S_1$ crosses with a triplet state every 30~fs on average.
Instead, the} main reason for the reduced ISC yield of 5-AC is likely the significantly smaller SOCs.
In 5-AC, the two states involved in ISC are the $^1n_\mathrm{N}\pi^*$ and $^3\pi\pi^*$ states; the SOC between those states is only 13~cm$^{-1}$.
On the contrary, in keto-cytosine, the two involved states are a $^1n_\mathrm{O}\pi^*$ and a $^3\pi\pi^*$; due to the localization of the transition density on the oxygen atom, the SOCs are increased to 40~cm$^{-1}$.
This increase of SOCs when going from an $^1n_\mathrm{N}\pi^*$ state to a $^1n_\mathrm{O}\pi^*$ state was previously reported for the keto and enol tautomers of cytosine.\cite{Mai2013C}
Because the ISC rate is approximately proportional to the square of the SOCs\cite{Marian2012WCMS} (if all other parameters are unchanged), the larger SOCs in cytosine can explain a tenfold increased ISC rate, compared to 5-AC.
It is worth mentioning here that also uracil and 6-azauracil show a $^1n_\mathrm{O}\pi^*$ state as their lowest singlet state, leading to the observation of significant ISC yields in both molecules.\cite{Hare2006JPCB,Kobayashi2009JPCA}


\section{Conclusions}

The photophysical behavior of 5-azacytosine after irradiation with UV light has been investigated by means of potential energy surface exploration and ab initio dynamics simulations including both singlet and triplet states.
The simulations were carried out with MRCIS(6,5)/def2-SVP electronic structure computations coupled to the surface hopping including arbitrary couplings method.
The MRCIS level of theory was carefully benchmarked against previously reported excited-state data obtained with the reliable CASPT2 method.

In our simulations, we found that the relaxation dynamics after UV irradiation of 5-azacytosine begins with the internal conversion from the bright $^1\pi\pi^*$ state to a dark $^1n_\mathrm{N}\pi^*$ state, localized on the N$_5$ atom which distinguishes 5-azacytosine from cytosine.
From the $^1n_\mathrm{N}\pi^*$ state, ground state relaxation proceeds with a time constant slightly larger than 1~picosecond, leading to ultrafast deactivation of this molecule.
The intersystem crossing yield has been estimated to $10\pm 8$\% based on the simulations, showing that intersystem crossing cannot compete with the significantly faster internal conversion to the ground state.

The differences between the dynamics of cytosine and the one of 5-azacytosine can be largely attributed to the different localization of the $n$ orbitals involved in their lowest singlet $n\pi^*$ state, as already suggested by Crespo and coworkers.\cite{Pollum2014TCC}
The $^1n_\mathrm{N}\pi^*$ state of 5-azacytosine both facilitates efficient ground state relaxation through the paths from its minimum to the $S_1/S_0$ CoIns and at the same time shuts down ISC by quenching the associated spin-orbit coupling matrix elements.


\section*{Acknowledgement}

S.\ M., P.\ M., and L.\ G.\ gratefully acknowledge funding from the Austrian Science Fund (FWF) within the project P25827, generous allocation of computer time on the Vienna Scientific Cluster (VSC3), and the European COST Action CM1204 ``XUV/X-ray light and fast ions for ultrafast chemistry'' (XLIC).
A.\ C.\ B. thanks the Conselho Nacional de Desenvolvimento Cient\'ifico e Tecnol\'ogico (CNPq) for the research fellowship and the Superintend\^encia de Tecnologia da Informa\c{c}\~ao (STI) of the University of S\~ao Paulo for services and computer time.



\providecommand*{\mcitethebibliography}{\thebibliography}
\csname @ifundefined\endcsname{endmcitethebibliography}
{\let\endmcitethebibliography\endthebibliography}{}

\end{document}